\documentstyle[epsfig,aps,prl,twocolumn]{revtex}
\newcommand{\etal}{{\em{et al.}}}
\newcommand{\Ptau}{P_\tau}
\newcommand {\BR}{{\mathrm{BR}}} 
\newcommand {\tnpp}{\tau^-\rightarrow\nu_\tau\pi^-\pi^0}

\newcommand {\tnnl}{\tau^-\rightarrow\nu_\tau \bar{\nu}_\ell \ell^-}
\newcommand {\MTval}     {(1777.05^{+0.29}_{-0.26})\mathrm{MeV}}             
\newcommand {\BRTRval}   {(25.31      \pm 0.18)        \%}                   
\newcommand {\GF}        {G_{\mathrm{F}}}    
\newcommand {\GFMUval}   {(1.16639  \pm 0.00001) \times 10^{-5}   \mathrm{GeV^{-2}}} 
\newcommand {\VUDval} {(0.9740 \pm 0.0010)}     

\newcommand {\KAPPAMval}  {0.001 \pm 0.008}               
\newcommand {\KAPPAEval}  {0.00  \pm 0.16}                
 
\newcommand {\KAPPAMlim}  {-0.014 < \kappa < 0.016}       
\newcommand {\KAPPAElim}  {|\tilde{\kappa}| < 0.26}       
\newcommand {\RHOKAPPAMval}  {0.16 \pm 0.08}                    
\newcommand {\RHOKAPPAEval}  {0.88^{+0.25}_{-0.35}}             

\newcommand {\RHOKAPPAMlim}  {0.00  < \kappa < 0.32}            
\newcommand {\RHOKAPPAElim}  {0.13 < |\tilde{\kappa}| < 1.33}   

\begin{document}
%
%
%
\draft
\title{Effects of anomalous charged current dipole 
       moments of the tau on the decay ${\mathbf{\tnpp}}$}
%
%
\author{Maria-Teresa Dova, Pablo Lacentre,}
\address{Universidad Nacional de La Plata, La Plata, Argentina}
\author{John Swain, and Lucas Taylor}
\address{Department of Physics, Northeastern University, Boston, USA}
\date{\today}
\maketitle
\begin{abstract}
We analyse the process $\tnpp$ allowing for anomalous weak charged 
current magnetic and electric dipole moment interactions and 
determine the effects on the differential and total decay rates.
Using recent experimental data we determine the following 
values for the anomalous magnetic and electric dipole moment 
parameters, respectively:
$$
 \kappa         = \RHOKAPPAMval;~~{\mathrm{and}}~~  \\
 |\tilde\kappa| = \RHOKAPPAEval. 
$$
\end{abstract}
\pacs{%
12.60.Cn, 
13.35.Dx, 
14.60.Fg, 
}
%
%
\section{Introduction}
In the Standard Model the charged current interactions of the tau lepton 
are mediated by the $W$ boson with a pure $V\!-\!A$ coupling.
We consider new derivative couplings in the Hamiltonian
which are parametrised by the parameters $\kappa$ and ${\tilde{\kappa}}$,
the (CP-conserving) magnetic and (CP-violating) electric dipole form 
factors respectively~\cite{RIZZO97A,CHIZHOV96A}.
These are the charged current analogues of the weak neutral 
current dipole moments, measured using $Z\rightarrow\tau^+\tau^-$ 
events~\cite{PICH97A}, and the electromagnetic 
dipole moments~\cite{BIEBEL96A,TTGNUCPHYSB}, measured using 
$Z\rightarrow\tau^+\tau^-\gamma$ events~\cite{OPALTTG,L3TTG,TAYLOR_TAU98}.
The only limits so far obtained for $\kappa$ and $\tilde{\kappa}$ 
are derived from analyses of the partial widths for 
$\tau^-\rightarrow \ell^- \bar{\nu}_{\ell}\nu_{\tau}$, 
for $\ell={\mathrm{e}}, \mu$~\cite{RIZZO97A,ANOMALOUS_COUPLINGS,DOVA_SWAIN_TAYLOR_TAU98}. 

In this paper we consider for the first time the effects of 
anomalous charged current dipole moments on tau decays involving hadrons.
We analyse the $\tnpp$ process which has largest branching fraction 
of all the tau decay modes.
This process is particularly topical due to a recently
reported difference of $2.2\sigma$ between the measured $\tnpp$ 
branching fraction and the (lower) value predicted using 
$e^+e^- \rightarrow \pi^+\pi^-$ data in the neighbourhood of the 
$\rho$ meson resonances and the Conserved Vector Current 
(CVC) hypothesis~\cite{TAU98_EIDELMAN}. 
While this could be attributed to a fluctuation, 
we note that non-zero values of $\kappa$ and $\tilde{\kappa}$ 
would also yield a higher measured value for $\BR(\tnpp)$.
 
We present predictions for the differential $\tnpp$ decay 
distributions and the partial width, $\Gamma(\tnpp)$, as functions of 
$\kappa$ and ${\tilde{\kappa}}$.
The sensitivity of the differential distributions is analysed for 
typical samples of $\tnpp$ decays in ${\mathrm{e^+e^-\rightarrow\tau^+\tau^-}}$
events which are reconstructed by the LEP and SLC experiments.
The partial width is compared to the experimental measurements of 
$\BR(\tnpp)$ to yield quantitative constraints on
$\kappa$ and ${\tilde{\kappa}}$.

\section{Parametrisation of anomalous couplings in ${\mathbf{\tnpp}}$ decays}

The matrix element for the decay $\tnpp$ is given by
\begin{equation}
 M = \frac {\GF}{\sqrt{2}} V_{ud} J^\mu H_\mu,
\end{equation}
where $\GF = \GFMUval$ is the Fermi constant,
$V_{ud} = \VUDval$ is the appropriate CKM matrix element~\cite{PDG98},
and $J^\mu$ and $H_\mu$ are the leptonic and hadronic currents respectively.
The effects of anomalous weak charged current dipole moment 
couplings at the $\tau\nu_\tau W$ vertex are parametrised by 
augmenting the usual $V-A$ charged current such that  
$J^\mu$ is given by
\begin{equation}
J^\mu   =  \bar{u}_\nu 
           \left( 
              \gamma^\mu \left( 1-\gamma^5 \right) - 
              \frac{i\sigma^{\mu\nu} q_\nu}{2m_\tau}  (\kappa - i\tilde{\kappa}\gamma_5)
           \right) u_\tau, \label{current}\\
\end{equation}
where $\sigma^{\mu\nu} = i /2[\gamma^\mu,\gamma^\nu]$, $q^\mu$ is the 
four-momentum transfer, and $m_\tau = \MTval$~\cite{PDG98} is the tau mass.
The parameters $\kappa$ and ${\tilde{\kappa}}$ are in general complex but henceforth 
we assume that ${\tilde{\kappa}}$ is real, as required by $CPT$ invariance.
The hadronic current is parametrised as 
\begin{equation}
H^\mu = \sqrt{2} F(q^2) (q_1 - q_2)^\mu,
\end{equation}
where $q_i$ denote the four-momenta of the two final-state \mbox{pions} and 
$F(q^2)$ is a form factor.

A convenient choice for the kinematic observables, following 
K{\"{u}}hn and Mirkes~\cite{KUHN92A}, is: 
$q^2$, the invariant mass-squared of the hadronic system; 
$\cos\theta$, the cosine of the angle between the tau spin-vector and 
the hadronic centre-of-mass direction as seen in the tau rest frame; 
and $\cos\beta$, the cosine of the angle between the charged pion
and the axis pointing in the direction of the laboratory viewed 
from the hadronic centre-of-mass frame (henceforth
referred to as the $z$-axis). 

\subsection{Differential Decay Rate, 
           ${\mathbf{d\Gamma(\tnpp)/dq^2 d\cos\theta d\cos\beta}}$}

After integration over the unobservable neutrino direction and the 
azimuthal angle of the charged pion, and neglecting the
mass difference between the charged and neutral pions, the differential 
decay rate is given by
\begin{eqnarray}
d\Gamma &  = & \frac{1}{(4\pi)^3} \frac{G_F^2}{4 m_\tau^3} 
                    |V_{ud}|^2 |F(q^2)|^2 S_{EW} \nonumber \\
        &   &  \times 
                    (q^2-4 m_\pi^2)^{3/2} (m_\tau^2-q^2)^2 \nonumber \\
        &   &  \times
                    \left\{ 
                       \left[ 
                          f_0+Re(\kappa) f_1+(|\kappa|^2+\tilde{\kappa}^2) f_2 
                       \right] 
                    \right.                                             \nonumber \\
        &   &       \left. 
                       \mbox{~~~}+ P_\tau 
                       \left[ 
                          g_0+Re(\kappa) g_1+\tilde{\kappa} Im(\kappa) g_2 
                       \right] 
                    \right\}                                             \nonumber \\
        &  &  \times \frac{dq^2}{\sqrt{q^2}} 
                     \frac{d\cos\beta}{2}  
                     \frac{d\cos\theta}{2}
                     \;,
\label{E1}
\end{eqnarray}
where $P_\tau$ is the tau polarisation and the factor of $S_{EW} = 1.0194$ 
accounts for electroweak corrections to leading logarithm~\cite{MARCIANO88A}. 
The functions $f_i$ and $g_i$ $(i=0,1,2)$ are given by
\begin{eqnarray}
f_0 & = & 2   \left[ 1+\frac{m_\tau^2-q^2}{q^2} Y \right], \\
f_1 & = & 1, \\
f_2 & = & 1/4   \left[ 1-\frac{m_\tau^2-q^2}{m_\tau^2} Y \right], \\
g_0 & = & 2   \left[ \frac{2 m_\tau}{\sqrt{q^2}} X - \cos\theta \right], \\
g_1 & = & \frac{m_\tau^2+q^2}{m_\tau \sqrt{q^2}} X - \left[ 1+\frac{(m_\tau^2-q^2)^2}{2 m_\tau^2 q^2} Y \right]   \cos\theta, \\
g_2 & = & \frac{\cos\theta}{2} - \frac{\sqrt{q^2}}{m_\tau} X,
\end{eqnarray}
where
\begin{eqnarray}
Y & =  & \frac{1}{3} \left[ 1+(3\cos^2\psi-1) \frac{3\cos^2\beta-1}{2} \right], \\
X & =  & \frac{m_\tau^2+q^2}{2 m_\tau \sqrt{q^2}} Y \cos\theta  
          + \sin\theta \frac{\sin 2\psi}{2} \frac{3\cos^2\beta-1}{2},
\end{eqnarray}
and $\psi$ is the angle between the tau direction of flight in the
hadronic rest frame and the $z$-axis. 
At LEP energies the following approximation is valid:
\begin{eqnarray}
\cos\psi & = & \frac{\eta + \cos\theta}{1+\eta \cos\theta},~~~{\mathrm{with}}     \\ 
\eta     & = & \frac{m_\tau^2-q^2}{m_\tau^2+q^2}.
\end{eqnarray}
$F(q^2)$ describes the resonant structure of the two-pion invariant mass
and the model used to describe it is discussed in more detail in the following
section.

\subsection{Dependence of Apparent Polarisation on Anomalous Couplings}

The usual determination of tau polarisation from energy and angular distributions
of decay products of the tau depends crucially on the assumed $V-A$ structure of
the charged current to serve as a polarimeter. 
Additional couplings in $\tau\rightarrow\rho\nu$ 
decays will produce measured values of polarisation which differ
from analyses of other $\tau$ decay modes and the 
predictions from global fits to Electroweak parameters in the context
of the Standard Model.
The observed agreement of polarisation measured in $\tnpp$ decays with
other determinations may be used to constrain $\kappa$ and $\tilde\kappa$.

We first integrate the differential width presented above with respect to $q^2$
and $\cos\beta$.
The $q^2$ dependence of $F$ must be explicitly considered prior to this integration.
$F(q^2)$ for the $\tnpp$ channel is dominated by the 
$\rho(770)$ vector meson with a small admixture of $\rho^\prime(1450)$
and a negligible contribution from the $\rho^{\prime\prime}(1700)$,
as verified by the ALEPH experiment~\cite{ALEPHRHOFIT}.
We work within the context of the K{\"{u}}hn and Santamaria model~\cite{KUHN90A}  
in which the $\rho$ and $\rho^\prime$ resonances are each described by a 
Breit-Wigner propagator with an energy-dependent width~\cite{DECKER93A}
\begin{equation}
B_x(q^2) = \frac{m_x^2}{m_x^2-q^2-i \sqrt{s} \Gamma_x(q^2)},
\end{equation}
where: 
\begin{eqnarray}
\Gamma_x(q^2) & = & \Gamma^0_x \frac{m_x^2}{q^2} \left( \frac{p(q^2)}
{p(m_x^2)} \right)^3; \qquad {\mathrm{and}}   \nonumber\\
       p(s)   & = & 1/2 \sqrt{s-4 m_\pi^2}. 
\end{eqnarray}
where $\Gamma^0_x$ is a constant.
The normalisation is fixed by chiral symmetry constraints in the limit of soft 
meson momenta, such that the form factor is given by
\begin{equation}
F(q^2) = \frac{B_\rho(q^2)+\beta B_{\rho^\prime}(q^2)}{1+\beta} \;.
\label{fks}
\end{equation}
where $\beta = -0.145$ \cite{KUHN90A}.
The differential width, retaining only the $\theta$ 
dependence, is of the form
\begin{equation}
d\Gamma = ( A + B P_\tau \cos\theta ) \frac{d\cos\theta}{2}.
\label{dcostheta}
\end{equation}
The coefficients $A$ and $B$ depend on $\kappa$ and 
$\tilde{\kappa}$ and are given by
\begin{eqnarray}
A & = & f^\dag_0+Re(\kappa) f^\dag_1+(|\kappa|^2+\tilde{\kappa}^2)
f^\dag_2 \nonumber\\
B & = & g^\dag_0+Re(\kappa) g^\dag_1+\tilde{\kappa} Im(\kappa) g^\dag_2
\end{eqnarray}
where, for $\beta = -0.145$, we obtain the following numerical values
\begin{eqnarray}
f^\dag_0 & = (518.6 \pm 6.4    ) \cdot 10^{-15}\,{\mathrm{GeV}}    \\
f^\dag_1 & = (111.3 \pm 1.9    ) \cdot 10^{-15}\,{\mathrm{GeV}}    \\
f^\dag_2 & = (20.74 \pm 0.38   ) \cdot 10^{-15}\,{\mathrm{GeV}}    \\
g^\dag_0 & = (221.6 \pm 2.04   ) \cdot 10^{-15}\,{\mathrm{GeV}}    \\
g^\dag_1 & = (-37.11 \pm .63   ) \cdot 10^{-15}\,{\mathrm{GeV}}    \\
g^\dag_2 & = (32.73 \pm 0.50   ) \cdot 10^{-15}\,{\mathrm{GeV}} 
\end{eqnarray}
Comparison of the results for the measured ``apparent polarisation'' from
$\tnpp$ channel with other channels or the results of Electroweak fits
would permit constraints to be placed on $\kappa$ and $\tilde{\kappa}$.

\subsection{Dependence of the Total Width, ${\mathbf{\Gamma(\tnpp)}}$ on
Anomalous Couplings}
Integration of Eq.~\ref{dcostheta} over $\cos\theta$ yields the 
effect of the anomalous couplings on the total rate 
\begin{equation}
\Gamma(\tnpp)  = \Gamma^0 
                  \left[1 + a_1 Re(\kappa) + 
                            a_2 (|\kappa|^2+\tilde{\kappa}^2) 
                  \right],
\label{gammarho}
\end{equation}
which naturally is independent of the polarisation 
(the polarisation term is proportional to $\cos\theta$ 
and therefore integrates to zero).
The parameters $a_1$ and $a_2$ are given by
\begin{eqnarray}
      a_1 & \equiv f_1^\dag/f_0^\dag & = 0.202; \qquad {\mathrm{and}}
              \\
      a_2 & \equiv f_2^\dag/f_0^\dag & = 0.037.
\end{eqnarray}
$\Gamma^0 (\equiv f_0^\dag)$ represents the partial width in the 
absence of anomalous couplings, {\em{i.e.}} $\kappa = \tilde{\kappa} = 0$.

\section{Sensitivity of the differential decay rate to $\kappa$ and ${\tilde{\kappa}}$}

The sensitivity of Eq.~\ref{E1} to the dipole moment couplings was 
studied for the case of CP-conserving interaction, i.e. $\kappa$ real and 
$\tilde{\kappa}=0$. 
We consider the quantity 
\begin{equation}
\sigma \sqrt{N} = \left[ \int d\Omega \frac{1}{f} 
                  \left(\frac{\partial f}{\partial \kappa}\right)^2\right]^{-1/2}
\end{equation}
as a function of the $\tau$ polarisation, where 
$\sigma$ is the expected error one standard deviation on $\kappa$, 
$N$ is the number of $\tnpp$ decays,
and $d\Omega$ is the elemental phase space volume.
The choice of the quantity $\sigma \sqrt{N}$ simply reflects 
the $1/\sqrt{N}$ dependence of the statistical error $\sigma$.

\begin{figure}[!htb]
\begin{center}
  \epsfig{file=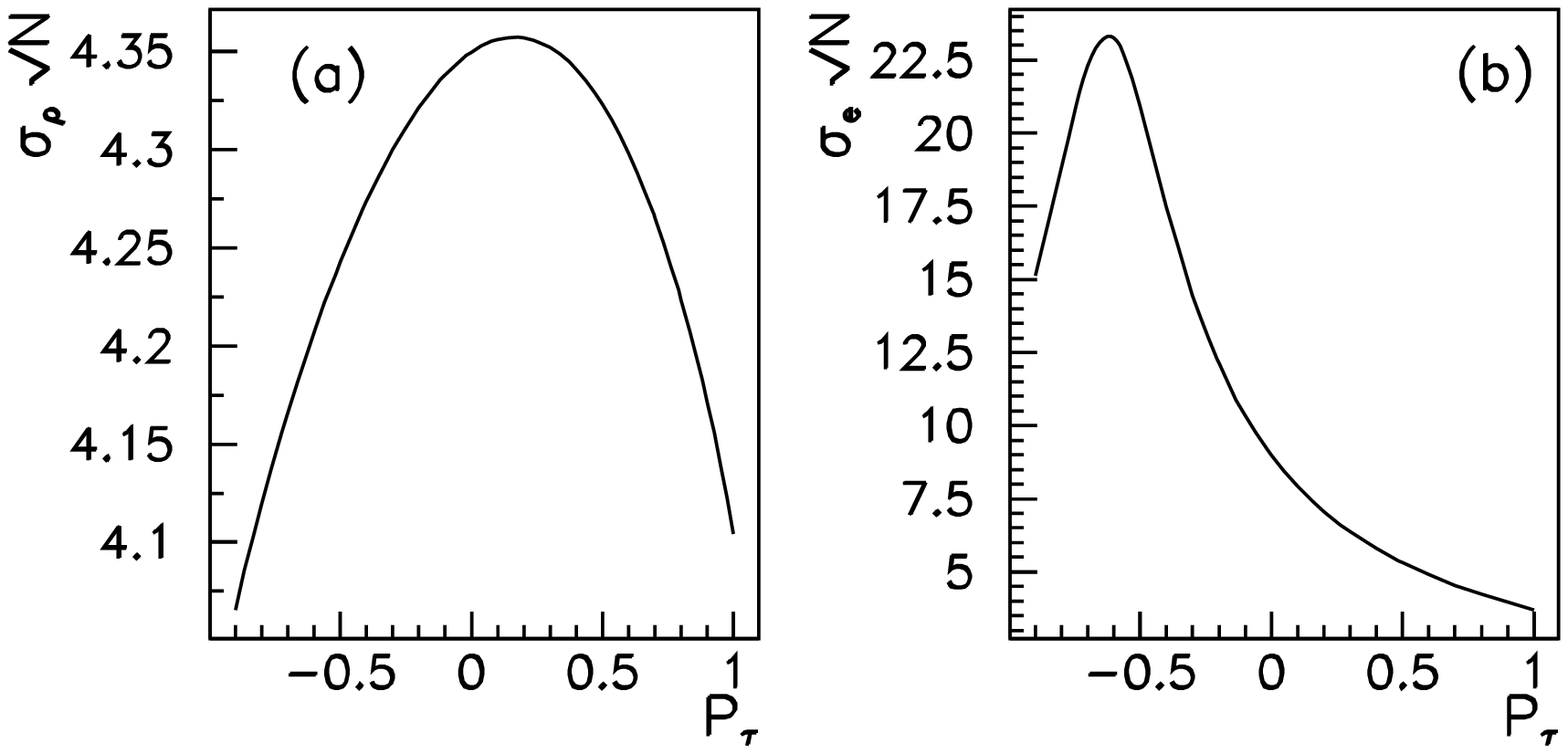,width=0.95\linewidth}
\end{center}
\caption{The $\kappa$ sensitivity quantity, $\sigma \sqrt{N}$, 
         as a function of $\Ptau$ for $\kappa\approx 0$,
         for (a) the $\tnpp$ decay mode and 
             (b) the $\tnnl$ decay mode.}
\label{F3}
\end{figure}

The distribution $f$ is given by Eq.~\ref{E1} for the 
$\tnpp$ channel and Eq.~11 of Rizzo~\cite{RIZZO97A} for 
$\tnnl$.
Figure~\ref{F3} shows $\sigma \sqrt{N}$ as a function of $\Ptau$ for the 
particular case $\kappa\approx 0$ for 
(a) the $\tnpp$ decay mode, and 
(b) the $\tnnl$ decay mode ($\ell = {\mathrm{e}}$ or $\mu$, not both combined). 
The $\tnpp$ mode is intrinsically more sensitive than the 
leptonic decay modes and is less dependent on
the $\tau$ polarisation. 
In addition, the branching fraction for $\tnpp$ is larger than  each 
leptonic channel.

For example, at the Z peak ($\Ptau \approx -0.15$) with a typical sample of 
reconstructed decays for each LEP experiment ($\sim 45\,000$ $\tnpp$ decays 
and $\sim 30\,000$ $\tnnl$ decays)
the predicted statistical errors are:
\begin{eqnarray}
\sigma_\rho    & \sim & 0.020 , {\mathrm{and}}    \\
\sigma_\ell    & \sim & 0.065 ,   
\end{eqnarray}
where detector effects are neglected, apart from their influence on
the reconstruction efficiency which is reflected in the number of 
decays assumed.
The corresponding statistical error for the combined e and $\mu$ channels is $\sim$0.046
which is more than a factor of two less precise than from the $\tnpp$ channel 
alone.

A practical disadvantage of the semileptonic decay is the multi-dimensional 
character of the distribution function. 
In the analysis of the tau polarisation, this problem has been overcome 
using a single ``optimal variable''~\cite{DAVIER93A}. 
Although the optimal variable for the tau polarisation is 
not the optimal variable for $\kappa$ (nor for $\tilde{\kappa}$), 
we find it still provides distinguishing power.
We fit hypothetical distributions of the optimal variable for 
simulated samples of 45\,000 $\tnpp$ decays each, generated with 
$\Ptau=-0.15$ to represent $\tau$'s produced at the Z peak.
Typical errors are $\sigma_\rho \sim 0.038$ which is degraded compared 
to the full multi-dimensional fit but is still statistically more sensitive 
than the combined e and $\mu$ channels.

The apparent disadvantage of the leptonic channels is, however, mitigated 
by the need to know $F(q^2)$ for the $\tnpp$ channel which has a non-negligible
systematic error, as discussed below.

In this paper we cannot derive constraints on $\kappa$ and ${\tilde{\kappa}}$  
from fits to the differential decay distributions due to a lack of the 
necessary experimental information.    
We can, however, determine constraints from the (intrinsically less sensitive) 
measurements of $\BR(\tnpp)$, as described below.

\section{Constraints on $\kappa$ and ${\tilde{\kappa}}$ from $\BR(\tnpp)$}
    
We derive quantitative constraints on $\kappa$ and ${\tilde{\kappa}}$ by considering the 
likelihood for the theoretical prediction for $\BR(\tnpp)$ 
to agree with the experimentally determined average 
value of 
\begin{equation}
\BR(\tnpp) = \BRTRval,
\label{brexpt}
\end{equation}
as a function of $\kappa$ and $\tilde{\kappa}$. 
$\Gamma^0$ of Eq.~\ref{gammarho} is determined from 
$e^+e^- \rightarrow \pi^+\pi^-$ data using CVC.
A combined analysis of all data, allowing for radiative corrections 
and $\rho-\omega$ interference, yields the CVC prediction of~\cite{TAU98_EIDELMAN}
\begin{equation}
\BR(\tnpp) = (24.52 \pm 0.33)\%,
\label{brcvc}
\end{equation}
where the error includes statistical and systematic uncertainties and 
conservatively allows for a possible systematic discrepancy of the 
DM1 data compared to CMD, CMD-2, and OLYA.
The experimental value of $\BR(\tnpp)$ is higher than the CVC prediction 
by $2.2$ standard deviations of the combined error.
 

We fix $\Gamma^0$ to the CVC prediction of $\BR(\tnpp)$ so that only 
$a_1$ and $a_2$ depend on the description of hadronic spectral function.
This reduces the sensitivity of the results to the details 
of the hadronic modelling.

\begin{figure}[!htb]
\begin{center}
  \epsfig{file=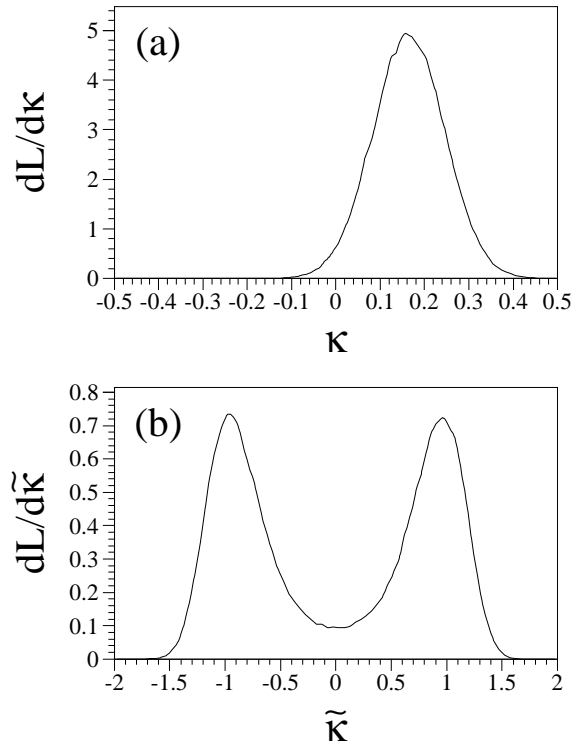,width=0.9\linewidth}
\end{center}
\caption{The likelihood distributions for 
         (a) $\kappa$ and (b) $\tilde{\kappa}$.}
\label{fig:probs}
\end{figure}

We construct likelihoods as a function of $\kappa$ and $\tilde{\kappa}$ 
conservatively assuming in each case that the other 
parameter is zero.
The errors are propagated numerically~\cite{NIM_LIKELIHOOD_PAPER} 
taking into account 
the error on the experimental measurement of $\BR(\tnpp)$ (Eq.~\ref{brexpt}),
the uncertainty on the CVC prediction (Eq.~\ref{brcvc}), 
a systematic error of $0.5$\% for radiative corrections not included 
in $S_{EW}$~\cite{MARCIANO88A},
and a systematic error of 0.3\% for the effect of $\rho-\omega$ 
interference~\cite{TAU96_EIDELMAN}.
Figure~\ref{fig:probs} shows the likelihood distributions for 
(a) $\kappa$ and (b) $\tilde{\kappa}$. 
The distribution for $\kappa$ has a single peak due to the dominance 
of the term linear in $\kappa$ in Eq.~\ref{gammarho}.
The distribution for $\tilde{\kappa}$ has two symmetric peaks due to 
the lack of a term linear in $\tilde{\kappa}$ in Eq.~\ref{gammarho},
therefore it is more appropriate to constrain the quantity 
$|\tilde\kappa|$.
We determine 
\begin{eqnarray}
  \kappa         & = & \RHOKAPPAMval ,~~~{\mathrm{and}}  \\
  |\tilde\kappa| & = & \RHOKAPPAEval ,
\end{eqnarray}
where the errors correspond to the 68\% confidence level.
At the 95\% confidence level we constrain $\kappa$ and $\tilde\kappa$ 
to the ranges:
\begin{eqnarray}
 & \RHOKAPPAMlim &;~~~{\mathrm{and}}  \\
 & \RHOKAPPAElim &.
\end{eqnarray}
These results are slightly more than two standard deviations from the 
SM expectations of zero which, though intriguing, cannot be 
considered statistically compelling evidence of new physics.

The results may be compared to complementary results previously 
obtained from purely leptonic tau decays~\cite{ANOMALOUS_COUPLINGS}, 
which are $\kappa = {\KAPPAMval}$ and $\tilde\kappa = {\KAPPAEval}$ or 
${\KAPPAMlim}$ and ${\KAPPAElim}$ at the 95\% C.L.~\cite{ANOMALOUS_COUPLINGS}.
These are more restrictive than those we obtain from $\tnpp$ decays.
This is partly due to the larger uncertainties in the theoretical 
and experimental values of the $\tnpp$ branching fractions.
In principle these effects could be counteracted by a higher intrinsic 
sensitivity of the $\tnpp$ channel due to larger values of $a_1$ and 
$a_2$ relative to the leptonic modes.
From our calculations, however, we see in retrospect that the numerical values 
for $a_1$ and $a_2$ are smaller than their leptonic counterparts 
(0.5 and 0.1 respectively).
Therefore, if only the total rate information is used
the $\tnpp$ channel is less sensitive than the leptonic channels,
in contrast to the higher statistical sensitivity of the $\tnpp$ channel 
when the differential decay distribution is analysed.

\section{Summary}

We present calculations of the differential and total decay rates 
for the process $\tnpp$, allowing for anomalous charged current 
magnetic and electric dipole moments, $\kappa$ and $\tilde\kappa$
respectively.
This constitutes the first such analysis of a hadronic 
tau decay mode.

The analysis of the differential distributions for the $\tnpp$ decay mode 
is found to be statistically more sensitive than the corresponding analysis 
of purely leptonic modes, $\tnnl$, irrespective of the tau polarisation.
The branching fraction, $\BR(\tnpp)$, is also sensitive to $\kappa$ and 
$\tilde\kappa$ although less so than for the leptonic branching fractions.

By comparison of the measured value of $\BR(\tnpp)$ with the predictions
of CVC, we determine
$\kappa         = \RHOKAPPAMval$ and
$|\tilde\kappa| = \RHOKAPPAEval$.
which differ from the SM expectations by approximately two 
standard deviations.

The values for $\kappa$ and $\tilde\kappa$ obtained 
from $\BR(\tnnl)$ are consistent with zero.
This could mean that the measured result for $\BR(\tnpp)$ and CVC 
differ only due to a fluctuation, or that there is a theoretical 
or experimental uncertainty which is not correctly 
taken into account.
The new $e^+e^- \rightarrow \pi^+\pi^-$ data in the 
neighbourhood of the $\rho$ meson resonances should 
reduce the experimental uncertainty in the CVC prediction 
by a factor of almost two in 1999~\cite{TAU98_EIDELMAN}.
Hopefully these data will clarify whether this is a statistical
or systematic effect or the first indication of some new physics
which affects hadronic tau decays but not purely leptonic decays.
 
\section*{Acknowledgements}
We acknowledge gratefully the financial support of CONICET
and the Fundaci{\'{o}}n Antorchas, 
Argentina (M.T.D. and P.L.) and the NSF, USA (J.S. and L.T.).
%



\end{document}